%% file: collate.tex
\magnification=1200

\input macros

\input title

\input intro

\input observ

\input linear

\input recons

\maybesupereject

\input conc

\begingroup \input refs \endgroup

\input ack

\input app

\closeout1 \vfill\eject
               \leftline{\bf\uppercase{figure captions}}
               \openout1=pscalls.tmp \input captions.tmp \closeout1
               \vfill\supereject \input pscalls.tmp \end

%% file: macros.tex


\input fonts

\input epsf

\expandafter
\ifx\csname Afour\endcsname\relax
\else
    \vsize=24.65truecm \hsize=15.94truecm
\fi

\expandafter
\ifx\csname draft\endcsname\relax
\else \input landscap
\fi

\def\psfigcall#1#2{\def\epsfsize##1##2{#2}\centerline{\epsfbox{#1}}}

\def\insfig[#1,#2,#3]#4.{\midinsert\parindent=0pt
           \vbox to #3 {\vss\psfigcall{#1}{#2}}
           \def\par{\endgraf\endinsert} \ninepoint
           \ignorespaces {\bf Figure#4.}\quad \ignorespaces}

\def\blackout{\catcode`\\=12\catcode`\{=12\catcode`\}=12\catcode`\$=12
\catcode`\&=12\catcode`\#=12\catcode`\^=12\catcode`\_=12\catcode`\@=12
\catcode`\~=12\catcode`\%=12\catcode`\|=12}

{\obeylines \global\let
                       \relax
    \gdef\fig{\begingroup\obeylines\verbatim}
    \gdef\verbatim{\blackout\def
                   ##1
                   {\setbox0\hbox{##1}%
                    \ifdim\wd0<1pt \endgroup \else\immediate\write1{##1}\fi
                   }
   \string\figcap}}

\def\figcap[#1,#2,#3]#4.{\write1{\null\vfil
           \vbox to #3{\vss\string\psfigcall{#1}{#2}}}
    \write1{\bigskip\bigskip\centerline{\twelvebf Figure#4}\vfil\string\eject}
    \bigskip\noindent{\bf\ignorespaces#4.}}

\expandafter
\ifx\csname submission\endcsname\relax
      \let\adjustlinespacing\relax
      
      \let\fig=\insfig
\else \def\adjustlinespacing{\baselineskip=1.5\baselineskip}
      \adjustlinespacing
      
      \openout1=captions.tmp
\fi

\def\maybesupereject{\expandafter\ifx\csname submission\endcsname\relax
                     \vfill\supereject\fi}
\catcode`\@=11

\newcount\notenum

\def\vfootnote#1{\insert\footins\bgroup\eightpoint
     \interlinepenalty=\interfootnotelinepenalty
     \splittopskip=\ht\strutbox \splitmaxdepth=\dp\strutbox
     \floatingpenalty=20000
     \leftskip=0pt \rightskip=0pt \parskip=1pt \spaceskip=0pt \xspaceskip=0pt
     \smallskip\textindent{#1}\footstrut\futurelet\next\fo@t}

\def\note{\global\advance\notenum by 1
    \edef\n@tenum{$^{\the\notenum}$}\let\@sf=\empty
    \ifhmode\edef\@sf{\spacefactor=\the\spacefactor}\/\fi
    \n@tenum\@sf\vfootnote{\n@tenum}}


\def\iterate{\let\next=\relax\body\let\next=\iterate\fi\next}

\def\mydef#1#2{\expandafter\ifx\csname#1\endcsname\relax
               \expandafter\def\csname#1\endcsname{#2}\else
               \errmessage{#1 is already \csname#1\endcsname}\fi}
\newif\ifneedrerun
\def\\#1{\expandafter\ifx\csname#1\endcsname\relax
         ??\needreruntrue \else \csname#1\endcsname\fi}

\def\parb{\par }
\openin1=names
\loop
\ifeof1\else
   \read1 to \cs
   \ifx\cs\parb\else \expandafter\mydef\cs \fi
\repeat

\immediate\openout2=names
\def\sdef#1#2{\expandafter\xdef\csname#1\endcsname{#2}%
              \immediate\write2{{#1}{#2}}}


\newcount\eqnum
\def\nextenum{\global\advance\eqnum by 1 \theeqnum}
\def\prevenum#1{{\advance\eqnum by -#1 \advance\eqnum by 1 \theeqnum}}
\def\nameeq#1{\sdef{#1}{\the\eqnum}\ignorespaces}
\def\putnum{\eqno(\nextenum)}
\def\theeqnum{\rm\the\eqnum}


\def\<#1>{{\left\langle#1\right\rangle}}

\def\witchbox#1#2#3{\hbox{$\mathchar"#1#2#3$}}
\def\leqsim{\mathrel{\rlap{\lower3pt\witchbox218}\raise2pt\witchbox13C}}
\def\geqsim{\mathrel{\rlap{\lower3pt\witchbox218}\raise2pt\witchbox13E}}

\let\twiddle=~ \def~{{}\ifmmode\widetilde\else\twiddle\fi}
\let\comma=\, \def\,{\ifmmode\comma\else\thinspace\fi}

\def\frac#1/#2{\mathchoice  {\hbox{$#1\over#2$}} {{#1\over#2}}
                {\scriptstyle{#1\over#2}}
                {#1\mskip-1.5mu/\mskip-1.5mu#2}}
\def\half{{\frac1/2}}

\def\pderiv(#1/#2){\mathchoice{\partial#1\over\partial#2}
    {\partial#1/\partial#2} {\partial#1/\partial#2} {\partial#1/\partial#2}}
\def\pderivp(#1/#2){\left(\pderiv(#1/#2)\right)}

\def\tderiv(#1/#2){\mathchoice{d#1\over d#2}{d#1/d#2}
                   {d#1/d#2}{d#1/d#2}}
\def\tderivp(#1/#2){\left(\tderiv(#1/#2)\right)}

\def\leftmatrix#1{\null\,\vcenter{\normalbaselines\m@th
    \ialign{$##$\hfil&&\quad$##$\hfil\crcr
      \mathstrut\crcr\noalign{\kern-\baselineskip}
      #1\crcr\mathstrut\crcr\noalign{\kern-\baselineskip}}}\,}

\def\rightmatrix#1{\null\,\vcenter{\normalbaselines\m@th
    \ialign{\hfil$##$&&\quad\hfil$##$\crcr
      \mathstrut\crcr\noalign{\kern-\baselineskip}
      #1\crcr\mathstrut\crcr\noalign{\kern-\baselineskip}}}\,}


\headline={\ifnum\pageno=0\hfil
           \else\ifodd\pageno\hfil\tenit\rhead\quad\hbox{\tenbf\folio}%
                \else\hbox{\tenbf\folio}\quad\tenit\lhead\hfil\fi
           \fi}

\footline={\hfil}

\parskip \smallskipamount

\def\maybebreak#1{\vskip 0pt plus #1\vsize \penalty -1000
                  \vskip 0pt plus -#1\vsize}

\outer\def\section#1\par{\maybebreak{.1}\bigskip\bigskip\bigskip
      \centerline{\eightrm\uppercase{#1}}
      \medskip\noindent}

\def\comment{$\{$\bgroup\eightpoint\it\aftergroup\endcomment\let\next}
\def\endcomment{$\}$}

\def\ordsuffix#1{\ifcase#1\or st\or nd\or rd\or th\or th\or 
    th\or th\or th\or th\or th\or th\or th\or th\or th\or 
    th\or th\or th\or th\or th\or th\or st\or nd\or rd\or 
    th\or th\or th\or th\or th\or th\or th\or st\fi}
\def\dayord#1{#1$^{\rm\ordsuffix{#1}}$}
\def\today{\expandafter\dayord{\number\day}
 \ifcase\month\or
 January\or February\or March\or April\or May\or June\or
 July\or August\or September\or October\or November\or December\fi
 \space \number\year}

\catcode`\@=12

\def\pseudocode{$$\vcenter\bgroup\dimen1=\parindent
        \advance\hsize by -2\dimen1
        \rightskip=\dimen1\parindent=0pt
        \def\begin{\par\advance\leftskip by \dimen1}
        \def\nigeb{\par\advance\leftskip by -\dimen1}
        \setbox1\hbox{$\langle\,$}
        \def\<{\hangindent\wd1\hangafter1$\langle\,$}
        \def\>{$\,\rangle$\par}}
\def\endcode{\egroup\eqno\vrule height0pt width.1pt depth0pt
                         \llap{(\nextenum)}$$}


\def\PG{{\ninerm PG}1115+080}  \def\pg{{\ninerm PG}1115}

\def\kmsmpc{$\rm km\,s^{-1}\,Mpc^{-1}$}

\def\Dl{D_{\rm L}}  \def\Ds{D_{\rm S}}  \def\Dls{D_{\rm LS}}
\def\Sigcrit{\Sigma_{\rm crit}}

\def\btheta{\hbox{\bmit\char"12}}
\def\balpha{\hbox{\bmit\char"0B}}
\def\bthes{\btheta_{\rm S}}
\def\thee{\theta_{\rm E}}

\def\Fl{F_{\rm lens}}

\def\Schech{{\ninerm S}97}  \def\Kri{{\ninerm K}93}
\def\BK{{\ninerm BK}97}

\def\indentpara{\bgroup\leftskip=\parindent\def\par{\endgraf\egroup}
                \noindent}

\def\lhead{}
\let\rhead=\lhead

%% file: fonts.tex
 at 10 truept
\font\fourteenrm=cmr10 scaled 1440
\font\twelvebf=cmbx10 scaled 1200

\font\ninebmit=cmmib10 at 9pt

\let\bmit=\tenbmit

\font\ninerm=cmr9
\font\ninei=cmmi9
\skewchar\ninei='177
\font\ninesy=cmsy9
\skewchar\ninesy='60
\font\nineit=cmti9
\font\ninesl=cmsl9
\font\ninebf=cmbx9
\font\ninett=cmtt9
\def\ninepoint{\textfont0=\ninerm \scriptfont0=\sevenrm 
              \def\rm{\fam0\ninerm}\relax
              \textfont1=\ninei \scriptfont1=\seveni 
              \def\mit{\fam1}\def\oldstyle{\fam1\ninei}\relax
              \textfont2=\ninesy \scriptfont2=\sevensy 
              \def\cal{\fam2}\relax
              \textfont3=\tenex \scriptfont3=\tenex 
              \def\it{\fam\itfam\nineit}\relax
              \textfont\itfam=\nineit
              \def\sl{\fam\slfam\ninesl}\relax
              \textfont\slfam=\ninesl
              \def\bf{\fam\bffam\ninebf}\relax
              \textfont\bffam=\ninebf \scriptfont\bffam=\sevenbf
              \def\tt{\fam\ttfam\ninett}\relax
              \textfont\ttfam=\ninett
              \setbox\strutbox=\hbox{\vrule
                   height8pt depth3pt width0pt}\baselineskip=11pt
              \let\bmit=\ninebmit
              \adjustlinespacing
              \rm}

\font\eightrm=cmr8
\font\eighti=cmmi8
\skewchar\eighti='177
\font\eightsy=cmsy8
\skewchar\eightsy='60
\font\eightit=cmti8
\font\eightsl=cmsl8
\font\eightbf=cmbx8
\font\eighttt=cmtt8
\def\eightpoint{\textfont0=\eightrm \scriptfont0=\fiverm 
                \def\rm{\fam0\eightrm}\relax
                \textfont1=\eighti \scriptfont1=\fivei 
                \def\mit{\fam1}\def\oldstyle{\fam1\eighti}\relax
                \textfont2=\eightsy \scriptfont2=\fivesy 
                \def\cal{\fam2}\relax
                \textfont3=\tenex \scriptfont3=\tenex 
                \def\it{\fam\itfam\eightit}\relax
                \textfont\itfam=\eightit
                \def\sl{\fam\slfam\eightsl}\relax
                \textfont\slfam=\eightsl
                \def\bf{\fam\bffam\eightbf}\relax
                \textfont\bffam=\eightbf \scriptfont\bffam=\fivebf
                \def\tt{\fam\ttfam\eighttt}\relax
                \textfont\ttfam=\eighttt
                \setbox\strutbox=\hbox{\vrule
                     height7pt depth2pt width0pt}\baselineskip=9pt
                \adjustlinespacing
                \rm}

%% file: title.tex
\pageno=0

\null

\vskip 2.5truecm

\font\fourteenrm=cmr10 scaled 1440
\centerline{\vbox{\hsize=11truecm
            \baselineskip=12pt\parindent=0pt\parskip=0pt
            \def\\{\hfil\break}
            \def\/{\vadjust{\vrule depth 0.5ex width 0pt}\\}
            \def~#1{{\ninerm#1}}
            \null\vfil
            {\fourteenrm Non-parametric reconstruction\/
             of the galaxy-lens in PG1115+080}
             \bigskip
             Prasenjit Saha\\
             Department of Physics (Astrophysics)\\
             Keble Road, Oxford ~O~X1 3~R~H
             \vskip1pt {\ninett saha@physics.ox.ac.uk}
	     \medskip
             Liliya L.R. Williams\\
             Institute of Astronomy\\
             Madingley Road, Cambridge ~C~B3 0~H~A
             \vskip1pt {\ninett llrw@ast.cam.ac.uk}
}}

\vfil

\noindent {\bf Abstract:} We describe a new, non-parametric, method
for reconstructing lensing mass distributions in multiple-image
systems, and apply it to \PG, for which time delays have
recently been measured.

It turns out that the image positions and the ratio of time delays 
between different pairs of images constrain the mass distribution in 
a linear fashion.  Since observational errors on image positions
and time delay ratios are constantly improving, we use these data 
as a rigid constraint in our modelling. In addition, we require the 
projected mass distributions to be inversion-symmetric and to have 
inward-pointing density gradients. With these realistic yet 
non-restrictive conditions it is very easy to produce mass 
distributions that fit the data precisely.

We then present models, for $H_0=42$, 63 and 84 \kmsmpc, that in each
case minimize mass-to-light variations while strictly obeying the lensing
constraints.  (Only a very rough light distribution is available at
present.)  All three values of $H_0$ are consistent with the lensing
data, but require quite different morphologies for the lensing galaxy.
If $H_0$ is low, the main lensing galaxy could be an early type.  If
$H_0$ is high, the galaxy is reconstructed as a late elliptical or
nearly edge-on disc; a binary or merging system is also plausible.

\vfil


\eject

\def\lhead{Non-parametric lens models}
\let\rhead=\lhead

%% file: intro.tex
\section 1. Introduction

Observational data permitting, it is always better to model 
physical systems non-parametrically, so as not to
impose our prejudices and preconceptions on the outside world.
In cases of limited observational constraints, a
non-parametric model is not an option: one needs a parametric
model to fill in the gaps left by our observational ignorance. 

To date, all models of multiply-imaged QSOs are parametric, i.e.,
based on a predefined model for the galaxy mass distribution,
like an isothermal sphere, de Vaucouleurs
profile, King profile, etc., (For example, see Kochanek 1991.) The
parameters of the model are determined by optimizing the fit
between the observables and the corresponding model-predicted values, 
like image positions. The reason for using parametric models is that 
the number of observables is usually small and there is no
need to go beyond parametric models. However, lately it has become 
apparent (Witt \& Mao 1997) that parametric elliptical lenses are not 
adequate for modeling quadruple-lens systems, i.e., no lens with well 
determined image positions is satisfactorily fit by an isolated 
elliptical galaxy lens.

The history of modeling of \PG\ is an illustrative example of
how lens modeling has evolved in the last 15 years or so.
The first published model of \pg\ (Young {\it et al.} 1981) 
considered a spiral galaxy-lens consisting of a highly flattened disk 
and a slightly flattened bulge. Since the two components had different 
velocity dispersions, scale lengths, and ellipticities, the model had 
six parameters 
even without the adjustable lens redshift. There were only six
observational constraints. The acceptability of this over-specified 
yet realistic model went to show that the image configuration could 
easily arise from gravitational lensing. Later modelers demonstrated 
that the image configuration could be achieved with less parameters; 
Narasimha {\it et al.} (1983) used a single component lens: a truncated, 
mildly elliptical King profile. Kochanek (1991) fit five different models, 
each with 5 free parameters. With the arrival of more observational
constraints, i.e. the galaxy position, more precise image positions 
(Kristian {\it et al.} 1993, hereafter \Kri, Courbin {\it et al.} 1997),
and time delays for the 
images (Schechter {\it et al.} 1997, hereafter \Schech), all the single 
galaxy parametric models became unacceptable (\Schech, 
Keeton \& Kochanek 1997). 

A very plausible explanation is that nearby external galaxies
contribute non-negligible shear, and possibly convergence at the 
location of the images. Adding external galaxies gives one at 
least two additional parameters, magnitude and position angle of the 
shear, and thus more freedom to fit the data. This approach has 
been explored by Witt \& Mao (1997), and Keeton \& Kochanek (1997).

A reasonable question to ask is, as the observational constraints 
tighten should one use more elaborate parametric models with more 
free parameters to get acceptable fits to the observations? How far 
can one justify this approach? 

Instead, we decided to explore non-parametric models of the galaxy 
mass distribution. In this paper we describe our modeling method 
and apply it to a quadruple gravitational lens system, \PG. 
We show that the lensing observational constraints in this system
(image positions and time delay ratio), combined 
with a few general, non-restrictive requirements for the galaxy mass 
distribution produce realistic galaxy models. 

Our original motive for pursuing the non-parametric approach was to 
place realistic limits on the value of $H_0$, as derived from the
time delays in the quadruple \pg\ system. 
It has been pointed out by many authors that $H_0$ determination
from a multiple-image lens system is preferable to the conventional 
distance ladder approach (Jacoby {\it et al.} 1992) since the former 
measures $H_0$ on truly cosmological scales, thereby avoiding any
mass inhomogeneities in the local Universe, and bypassing all
the rungs in the distance ladder with their respective uncertainties.
The lensing method is not without its own problems; the derived
value of $H_0$ is sensitive to the uncertain mass distribution in 
the lens. Thus far only parametric mass models of \pg\ have been 
considered in the literature. However, there is always a worry 
with parametric models that only a small piece of the whole allowable 
model space is being looked at, so errors in the derived mass model,
and consequently errors in $H_0$ would be underestimated. 
In fact, our modeling 
clearly shows that the current parametric models do not cover all the
possible realistic galaxy mass distributions, since our derived galaxy
masses look different from the existing parametric models.
An example of models that are realistic but are not allowed
by the existing single-galaxy parametric models are mass 
distributions with twisting iso-density contours. 




%% file: observ.tex
\section 2. Observations of PG{\ninerm 1115+080}

\PG\ was the second multiply-imaged QSO to be discovered
(Weymann {\it et al.} 1980). It was originally selected as a QSO 
candidate based on its UV excess, and later spectroscopically 
confirmed and entered into the bright Palomar-Green QSO sample
(Green, Schmidt, Liebert 1986). Its redshift, based on several
emission lines, is $z=1.722$; its magnitude is $V=15.8$.
The multiple nature of \pg\ first became apparent during the
observations of Weymann {\it et al.}, when 
during brief periods of good seeing images B and C
were resolved by the TV guiding monitor at the 2.3-m at Steward
Observatory. Soon after, image A was shown to consist of two
images 0.5 arcseconds apart (Hege {\it et al.} 1981).

Several imaging observations were carried out on \pg, but to date, the
best astrometry on the system is from WFPC HST observations reported
by \Kri, and from NOT and CFHT observations reported by Courbin {\it
et al.} (1997). We will use the positions from K93 in this paper, but
results using the Courbin {\it et al.} astrometry are nearly
identical.

Even though it was recognized early on that the geometry of image
configuration can be reproduced by a single lensing galaxy (Young {\it
et al.} 1981), the identity and location of the galaxy remained
elusive for several years. The galaxy was eventually identified by
Shaklan \& Hege (1986), Christian {\it et al.} (1987), and \Kri. It
turned out to be a rather red faint object, $R\sim 20$, and $V-R\sim
0.7$, situated between the QSO images, and somewhat closer to B. In
addition to the main lensing galaxy, a nearby group of 3 galaxies was
found about 10--20 arcseconds south-west of \pg\ (Young {\it et al.}
1981, Henry \& Heasley 1986).  Henry \& Heasley (1986) obtained the
redshifts of the two brighter galaxies, $z=0.305$; while
Angonin-Willaime {\it et al.} (1993) determined the redshift of the
main lensing galaxy to be $0.295\pm0.005$. Just recently, Kundi\'c
{\it et al.} (1997) and Tonry (1997) obtained Keck spectra for all
four group galaxies, and showed that they are all at $z=0.311$, and
belong to a group with a line-of-sight velocity dispersion of $270\pm
40$~km~s$^{-1}$, and properties similar to those of Hickson's compact
groups.  Here, we will assume that the lensing galaxy is at $z=0.311$,
and its position is as given in \Kri.  As yet the lensing galaxy's
morphological type is unknown, but it is probably an elliptical
because (1)~the image positions very robustly require
$\sim10^{11}h^{-1}M_\odot$ within a radius of $3h^{-1}\rm\,kpc$;
(2)~the galaxy spectrum (Kundi\'c {\it et al.} 1997) shows no emission
lines that would be indicative of a merger or ongoing star-formation;
(3)~Tonry (1997) measures the velocity dispersion of the galaxy to be
281--293$\rm km\,s^{-1}$, depending on the aperture correction.

Unlike the 2-image QSO 0957+561, there is probably no other 
significant mass concentrations along the line of sight to \pg. 
Based on IUE data, Tripp {\it et al.}
(1990) report that there is no high column density Lyman-$\alpha$
absorption in the QSO spectrum, above the redshift of 0.39.
Thus we assume that all the mass relevant to lensing calculations
is in a single screen at $z=0.311$.

There have been a number of reports of image variability in this
system (Vanderriest {\it et al.} 1986, Foy {\it et al.} 1985, 
Michalitsianos {\it et al.} 1996). These observations were done 
under varying conditions, by a number of observers using different 
filters and instruments, and with large time lags between data points. 
As was pointed out by \Kri, the heterogeneous nature of accumulated 
data precludes one from making any positive conclusion regarding 
image variability either between images, or between same images at 
different epochs. 

Recently, \Schech\ obtained a
set of homogeneous photometric data on the A1+A2, B, and C images of
\pg\ on 30 nights that span a period of about 200 days in
1995-1996. The noise level varies from 4 millimag for combined images
A1+A2, to 13 millimag for image B. The lightcurves show systematic
variations of sufficiently large amplitude, $\Delta m\sim 0.15$ mag,
to derive relative time delays between images. A preliminary
evaluation of time delays by \Schech\ was soon followed by another
analysis of the same data by Bar-Kana (1997; hereafter \BK), who
incorporated the possible effects of correlated errors and
microlensing. \BK\ estimates that B lags C by $25^{+3.3}_{-3.8}$ days,
consistent with the \Schech's $23.7\pm3.4$ days. The delay between B 
and A is less certain, $\sim$10--14 days.  The ratio between time delays
consequently remains somewhat uncertain; \BK\ estimates $t_{\rm
AC}/t_{\rm BA}=1.13^{+0.18}_{-0.17}$ compared with \Schech's
$\simeq0.6$.  We will use the \Schech\ values for one reconstruction in
this paper, but mostly we will work with the \BK\ values.

Since the images are superimposed on the central part of the lensing
galaxy, the optical image fluxes probably suffer from microlensing. 
This precludes one from using the optical flux ratios as a modeling
constraint. Radio emission from a QSO would not be affected by 
microlensing, but \pg\ is radio-quiet, i.e. it is not detected down to 
15 mJy level with the VLA (Weymann {\it et al.} 1980). 
However, even if microlensing-free radio flux ratios existed,
they would not provide a good model constraint: between the time
delays, image positions and image amplifications, the latter are
the most sensitive to small changes in the lensing model. This is 
because image fluxes are a function of the second derivative of the 
2D lensing potential, whereas the time delays and image positions 
are proportional to the lensing potential and its first derivative, 
respectively. (See also Section 3.) We do not use image fluxes ratios
as constraints in our modeling.

To sum up, \PG\ is a quadruply-lensed bright optical QSO, with well
defined image positions (5 mas errorbars or better), a dominant 
lensing galaxy (50 mas position errorbars or better), and well constrained 
and easily observable time delays of $\sim$ several days to a couple 
of weeks. 

Before we proceed we would like to emphasize that not all observational 
constraints are equally good for the purposes of lens modeling, even if
the observational errors are small. The best observables are the ones
that depend on the global properties of the lens, with little sensitivity
to local mass variations. As we already mentioned, image fluxes are not 
reliable for that reason, while image positions are more stable. 
Time delays between images are also very good, but since time delays
themselves involve $H_0$, we use the {\it ratio} of the two observed
time delays, $t_{AC}/t_{BA}$. Thus we have 9 constraints: 8 from image
positions, plus the time delay ratio. In principle, the errors on these
quantities will keep on shrinking with improving data. Therefore, we
use these data as rigid constraints.

%% file: linear.tex
\section 3. A lens reconstruction method

Here we describe our method of non-parametrically reconstructing the
galaxy mass distribution based on observed lensing constraints, and
a few general requirements for the galaxy.
We take the lens as a pixellated mass distribution.  It is simplest to
leave masses in terms of the critical density, so we say that the
$n$th pixel has surface density $\kappa_n$ times
$$ \Sigcrit = {\Dl\Ds\over\Dls} {c^2\over 4\pi G},  \putnum$$
$\Sigcrit$ being the critical surface mass density for lensing
in units of mass per unit solid angle at the redshift of the lens.
The $n$th pixel's contribution to the potential at some $\btheta$ is
$\kappa_n\psi_n(\btheta)$ and its contribution to the bending angle is
$\kappa_n\balpha_n(\btheta)$, where
$$ \eqalign{&\psi_n(\btheta) = {1\over\pi} \int\! \ln|\btheta-\btheta'| \,
                   d^2\btheta', \cr
            &\balpha_n(\btheta) = {1\over\pi} \int\!
                   {\btheta-\btheta'\over|\btheta-\btheta'|^2} \,
                    d^2\btheta'. \cr }   \putnum$$
\nameeq{pixelintegs}
The integrals in (\\{pixelintegs}) are over the $n$-th pixel, and
depend on where $\btheta$ is relative to that pixel, and on the pixel
size (say $a$).  Explicit expressions $\psi_n$ and $\balpha_n$
for square pixels are worked out in the Appendix.

To get the full time delay and bending angle we sum over all pixels.
For a source at $\bthes$ the time delay is proportional to
$$ \tau(\btheta) = \half|\btheta|^2 - \btheta\cdot\bthes -
   \sum_n \kappa_n\psi_n(\btheta).  \putnum$$
\nameeq{delayeqn}
(Compared to how (\\{delayeqn}) is usually written, we have dropped
the term ${1\over2}|\bthes|^2$ because it cancels for all observable 
effects.)
To get the time delay in physical units we have to multiply $\tau$ by
$$ (1+z_{\rm L}) {\Dl\Ds\over c\Dls}. \putnum$$
\nameeq{timescale}
Images will appear where $\nabla\tau(\btheta)=0$, or
$$ \btheta - \bthes - \sum_n \kappa_n\balpha_n(\btheta) = 0.  \putnum$$
\nameeq{poseqn}

Now, consider what lensing data imply when inserted into equations
(\\{delayeqn}) and (\\{poseqn}).  First, the image positions in a
multiple-image system tell us that $\nabla\tau(\btheta_i) = 0$ for
some known $\btheta_i$.  Each image provides two such constraint
equations.  From (\\{poseqn}) we can see that these constraint
equations are linear in the unknowns ($\kappa_n$ and $\bthes$).
Secondly, if time delays have been measured between more than one pair
of images, as in \Schech, then the ratio of time delays provides a
constraint of the type
$$ \tau(\btheta_4)-\tau(\btheta_1) =
   \<\rm ratio> \times (\tau(\btheta_2)-\tau(\btheta_1)).  \putnum$$
Comparing with (\\{delayeqn}), we see that such constraints are also
linear.

Thus the lensing data on \PG\ provide nine constraint
equations, eight from the image positions and one from the time delay
ratio.  We will supplement these with some more constraints on
physical grounds.
\item{(i)} Obviously, the $\kappa_n$ must be non-negative.
\item{(ii)} The centroid of the lensing galaxy is known from \Kri, and
we require the mass profile to be $180^\circ$ rotation symmetric about
this centroid.  In other words $\kappa_{i,j}=\kappa_{-i,-j}$, where we
are changing notation briefly from $\kappa_n$ to $\kappa_{i,j}$ and
$\kappa_{0,0}$ is centred on the galaxy centroid.  We remark that
without this constraint or some other that fixes the origin, the
lensing data would have supplied only seven constraint equations.
\item{(iii)} For a galaxy we expect the density gradient to generally
be pointing inwards.  Accordingly, we define a pixellated density
gradient
$$ \nabla \kappa_{i,j} \equiv (2a)^{-1}
   \left(\kappa_{i+1,j}-\kappa_{i-1,j}\,,\,
   \kappa_{i,j+1}-\kappa_{i,j-1}\right)   \putnum$$
\nameeq{densgrad}
and require that $\nabla \kappa_{i,j}$ point no more than $45^\circ$
away from radially inwards.
\item{(iv)} We require that $\tau({\rm A1})<\tau({\rm A2})$, i.e., of
the two A images, the southern one leads.\note{Given that C leads A
leads B, there are two natural five-image configurations. One of them
has A1, A2 and C as respectively minimum, saddle-point and mininum in
a lemniscate, and B and the demagnified central image being
respectively saddle-point and maximum in a lima\c con.  The other
configuration has A1, A2 and C as respectively saddle-point, maximum
and mininum in a lima\c con.  (See Figure~6 in Blandford and Narayan
1986.) Both of these have A1 leading A2, and in fact we have only ever
seen the former in models of the \pg\ system.  Other configurations
are logically possible, but would have contorted time delay surfaces.}
\item{(v)} Though one can generate a mass profile and then calculate
the implied $H_0$, in practice we found it more convenient to set
$H_0$ at various values and generate different mass profiles at each
value. Since the time scale (\\{timescale}) is $\propto H_0^{-1}$,
fixing $H_0$ is just another linear constraint.

The foregoing are all linear equality and inequality constraints and
it is straightforward to find, by linear programming, if there are any
feasible mass distributions.  If there are then there is not one mass
model but a huge family of them, all consistent with the lensing data
and our additional imposed constraints.  At this point there are
several sensible things one can do.  One might take some sort of
average over all the allowed mass distributions.  Or one might specify
some figure of merit such as smoothness or entropy, and then optimize
subject to all the above constraints.  In this paper we will do the
latter, choosing our optimal model as the one with minimum 
mass-to-light variation.  More precisely, given a pixellated light
distribution $L_n$ for the galaxy, we find the mass distribution that
minimizes
$$ \sum_n(\kappa_n-KL_n)^2, \quad {\rm where\ }
   {\textstyle K=\sum_n\kappa_n, \ \sum L_n=1,}   \putnum$$
subject to all the linear constraints above.

Our reconstruction procedure has a simple geometrical
interpretation. Supposing there are $N$ independent pixels, consider
the $N$-dimensional space of the independent $\kappa_n$.  The lensing
constraints and $H_0$ specification confine allowed models to a
$(N-8)$-dimensional hyperplane in this space.  (There are ten equality
constraints in all, but two of them go to solve for $\bthes$.)  The
inequality constraints further confine allowed models to a convex
region of the hyperplane.  Now $L_n$ is a point in the $N$-dimensional
space, and a line from the origin through $L_n$ represents constant
mass-to-light models.  Our optimal model is then the point on the
allowed part of the $(N-8)$-hyperplane that is closest to that line.

The numerical problem has a unique solution, and quadratic programming
algorithms reliably find it.  We used the {\ninerm NAG} routine
{\ninerm E}04{\ninerm NFF}.

In this paper we do not use flux ratios as additional constraints,
because they are sensitive to local variations of $\kappa$ rather than
global variations and are thus weakly coupled to time delays, as was
explained in Section 2.  Therefore we did not attempt to constrain
local variations of $\kappa$ in our mass reconstructions, and in fact
our mass maps have unphysically large pixel-to-pixel variation.  This
pixel-to-pixel noise does not matter for image positions and time
delays because the potential remains very smooth (see Section~4
below), but it makes the magnification unphysically noisy.  In
principle, flux ratio fitting could be introduced into the
reconstruction method in a straightforward way. Since inverse
magnification involves products of the derivatives of the lens
equation (\\{poseqn}) it will be quadratic in the unknowns, and hence
constraints involving magnification could be dealt with by the same
numerical algorithms.  But to avoid a situation where flux ratios just
get fitted with pixel-to-pixel noise, it would be necessary to
suppress the latter using suitable physically motivated constraints.

Is pixellating the mass distribution the best approach to
non-parametric reconstruction?  We don't know. Other strategies using
either basis functions or grids with interpolation are certainly
possible; these would avoid discontinuities in the mass distribution,
and small scale inhomogeneities will be easier to suppress, but
non-negativity may be much harder to enforce.  But pixels have a
certain conceptual simplicity, and mainly for this reason we decided
that pixels should at least be investigated first.

%% file: recons.tex
\section 4. Reconstructions of the lens

For the \pg\ system, the time scale factor (\\{timescale}) equals
$31h^{-1}\rm\,days\,arcsec^{-2}$ for $q_0=\half$.  The $q_0$
dependence is weak and is easily read off Figure~1, so in the text we
will write ``$h=0.5$'' as shorthand for ``$H_0=50$ \kmsmpc for
$q_0=\half$'' and so on.  The angular diameter distance of the lens is
$\Dl=2.77h^{-1}\rm\,kpc\;arcsec^{-1}$ and the critical density
$\Sigcrit=3.30h^{-1}\times10^{10}M_\odot\rm\,arcsec^{-2}$. The total mass
in a pixellated reconstruction is
$$ \Sigcrit \times {\textstyle a^2\sum_n\kappa_n}.  \putnum$$
We can define a formal Einstein radius $\thee$ by equating
$a^2\sum_n\kappa_n$ to $\pi\thee^2$.  (It is only formal because the
lensing mass isn't circular and doesn't produce Einstein rings.)  Then
the total mass is
$$ \thee^2\times 1.04\times 10^{11}h^{-1}M_\odot\,\rm\,arcsec^{-2}.  \putnum$$
\nameeq{formal_Ering}
We can use equation (\\{formal_Ering}) to associate formal Einstein rings
with the other group galaxies too; it's a convenient way to show
various masses on a figure.
\fig[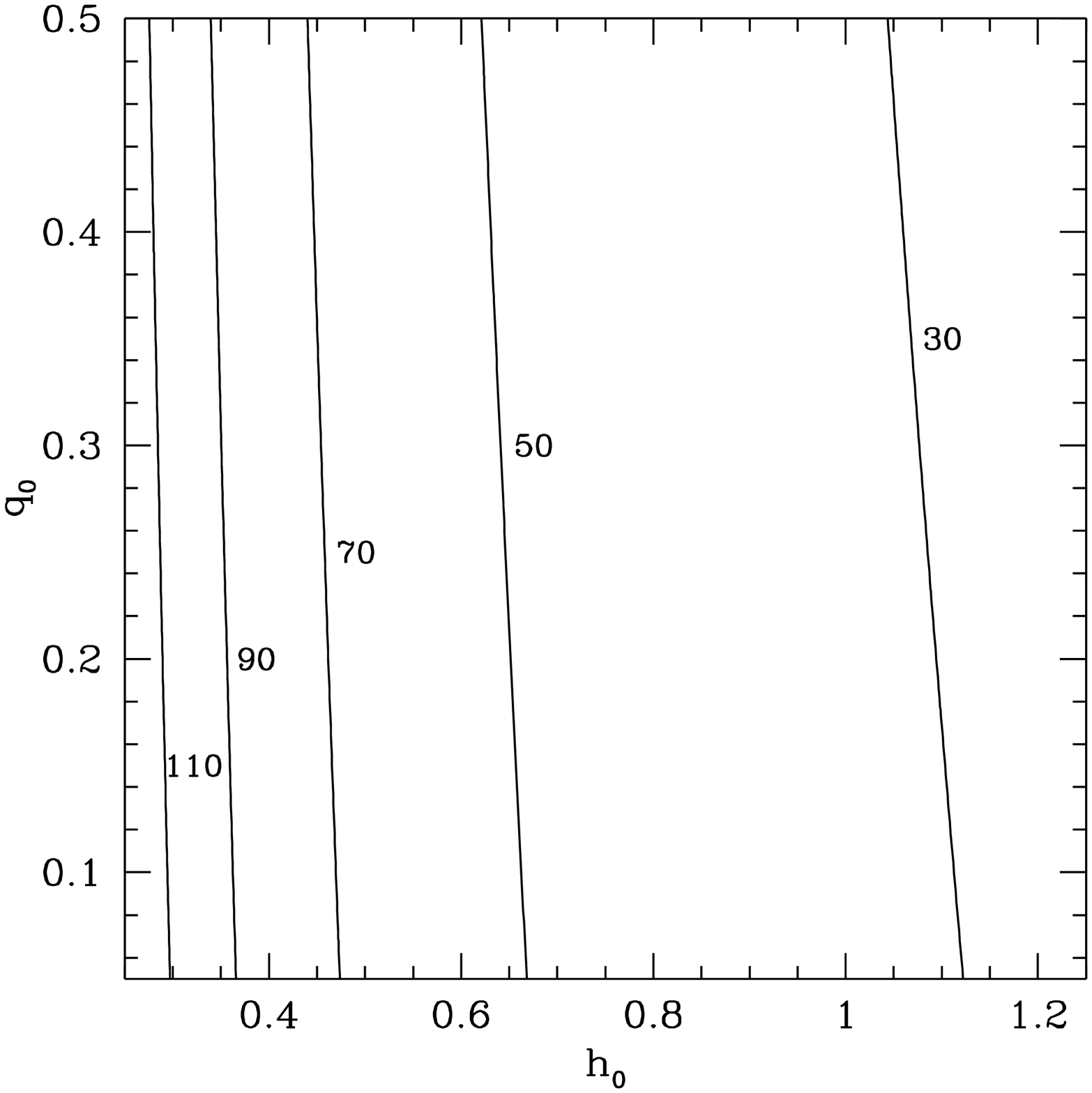,.7\hsize,.7\hsize] 1. Contours of constant time delay in
the $q_0$ vs. $h_0$ plane, expressed as the time factor given by
equation (\\{timescale}). Contour lines are labelled in units of
days~arcsec$^{-2}$.

In this section, we illustrate four reconstructions in detail. All
have $0.1''$ mass pixels and confine the main lensing galaxy to a
circle of radius $2''$ around the centroid measured by \Kri.  (That
makes 1265 pixels, but only 633 are independent because of the
$180^\circ$ rotation symmetry condition.) The other group galaxies, if
they are included, are considered as three point masses.  We did not
find our reconstructions sensitive to the pixel size and maximum
galaxy radius.  For the light distribution against which we optimize
we used the \Kri\ model
$$ L(\btheta) =
\left(1+{|\btheta|^2\over\theta_0^2}\right)^{-\gamma/2},
\quad \theta_0=0.71'',\ \gamma=1.7.  \putnum$$
\nameeq{lightmodel}
Again, our reconstructions are not very sensitive to the precise
values of $\theta_0$ and $\gamma$.

\fig[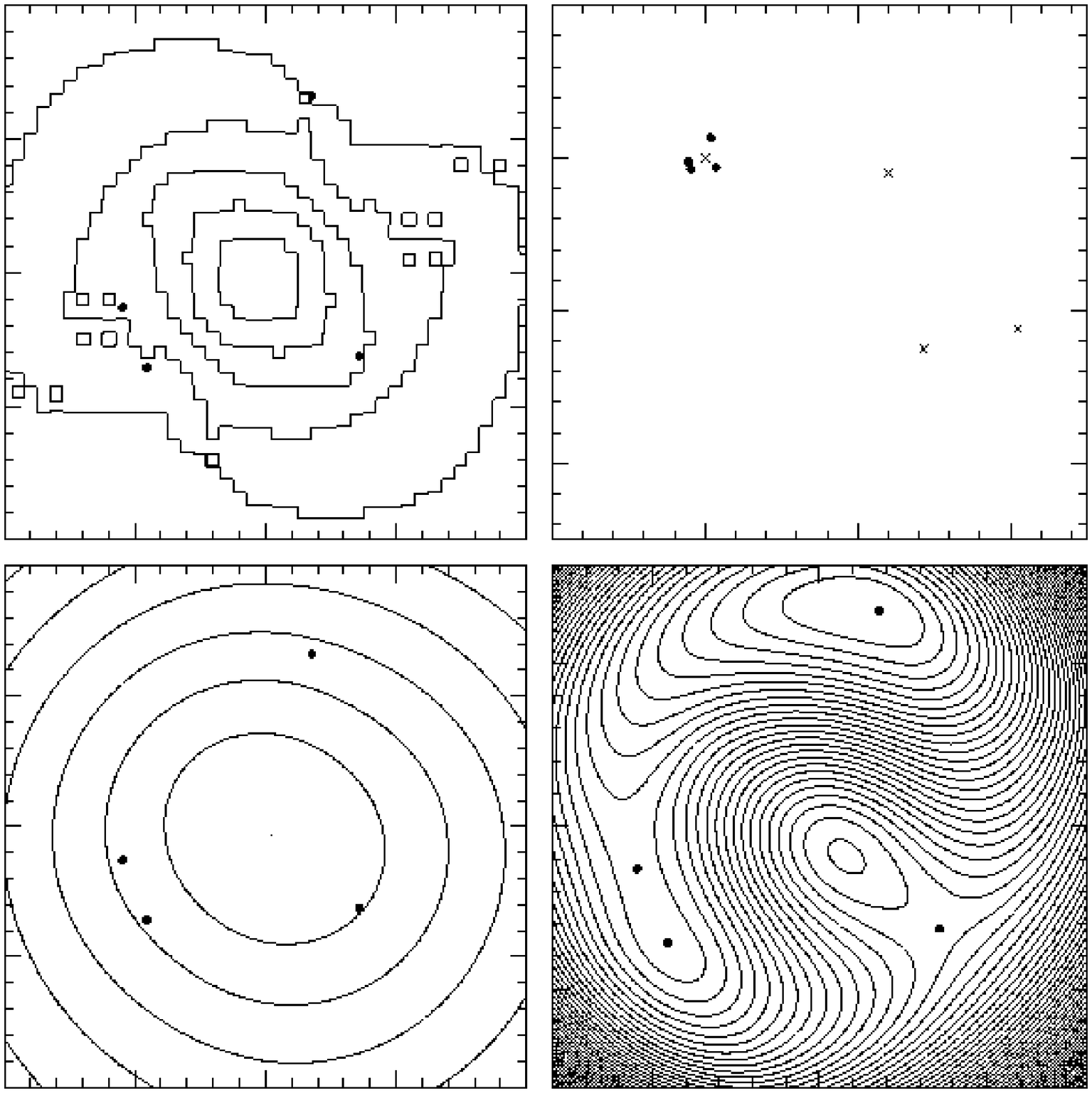,\hsize,\hsize] 2. A reconstruction constrained by
the image positions from \Kri\ and the time delays from the \Schech\
analysis, and by $h=0.5$.
\endgraf
We have suppressed axis labels to reduce clutter, but the image
positions (marked by filled circles) in each panel should make the
scales clear.
\endgraf
{\bf Upper left panel:} Contours of the dimensionless surface density
$\kappa=\frac1/3,\frac2/3,\ldots$, increasing inwards.  The field
shown is $(-2'',-2'')$ to $(2'',2'')$ with the galaxy centroid from
\Kri\ at the origin.
\endgraf
{\bf Lower left panel:} Contours of
$\btheta\cdot\bthes+\psi(\btheta)$, increasing outwards in steps of
$\half\rm\,arcsec^2$.  The field is as in the upper panel.
\endgraf
{\bf Lower right panel:} Time delay contours, in steps of 1~day.  The
field shown is $(-1.6'',-1.6'')$ to $(1.6'',1.6'')$ with the same
origin.
\endgraf
{\bf Upper right panel:} The field shown here is $(-10'',25'')$ to
$(-25'',10'')$ relative to the same origin as before.  The crosses
mark the main lensing galaxy and the other group galaxies; the latter
are not included in this reconstruction, but will be in later ones.

Figure~2 shows a minimum mass-to-light variation model that fits the
astrometry from \Kri\ and the time delays from \Schech\ (C leads B by
23.7 days and A by 9.4 days) with $h=0.5$.  (This model does not
include the other group galaxies, but their positions are indicated in
the figure.)  \Schech\ fitted several parametric models to these data,
finding that unless shear from the external galaxies was included the
fits to the image positions were very poor, while the predicted time
delay ratio was inconsistent ($\simeq1.5$ versus $\simeq0.6$) even
with external shear.  Keeton \& Kochanek (1997) subsequently fitted
more parametric models, with similar results.  But for non-parametric
models such difficulties evaporate; it is easy to fit all the lensing
data precisely, even without external shear.

Though the model in Figure~2 is not one of the main results of this
paper (because the external galaxies have been omitted) it will do to
illustrate the reconstruction method, so let us discuss the figure in
some detail.

\indentpara
{\sl Density map:\/} The $180^\circ$ rotation symmetry allows
isodensity contours to twist, and this seems to be a generic feature of
our reconstructions though it is not always as pronounced as in this
model.  Some parametric models get a similar effect by using two shear
axes.  (Incidentally, the single-pixel peaks in the density map may
seem inconsistent with our density gradient constraint. But because
$\nabla\kappa_{ij}$ is formulated as a symmetric difference in equation
\\{densgrad}, such single-pixel blemishes are in fact tolerated by the
constraint.)

\indentpara
{\sl Lens potential:\/} We plot $\bthes\cdot\btheta+\psi(\btheta)$
rather than $\psi(\btheta)$ itself.  Comparing with models that
include the external galaxies, we have found that both $\bthes$ and
$\psi(\btheta)$ change drastically on including those galaxies, but
$\bthes\cdot\btheta+\psi(\btheta)$ (like the time delay surface)
changes very little.  In a sense there is a degeneracy between
$\bthes$ and the mass distribution, and in the plot we have used it to
nominally move the source to the origin.  It is remarkable how much
smoother the potential is than the density; of course the smoothness
comes from having integrated the density twice with a
$\ln|\btheta-\btheta'|$ kernel.

\indentpara
{\sl Time delay contours:\/} We have spaced the contours by 1~day, and
this plot provides a pleasing visual check that our fitting program
does what it is supposed to do.  From this plot the central image may
seem worryingly bright, but this is not really an issue.  The
reconstructions in this paper all have wide cores simply because the
simple \Kri\ light model does.  But we could imagine reshaping the
core in a circularly symmetric way to have a central density cusp;
this would demagnify the central image without affecting the positions
or time delays of the other images.

\fig[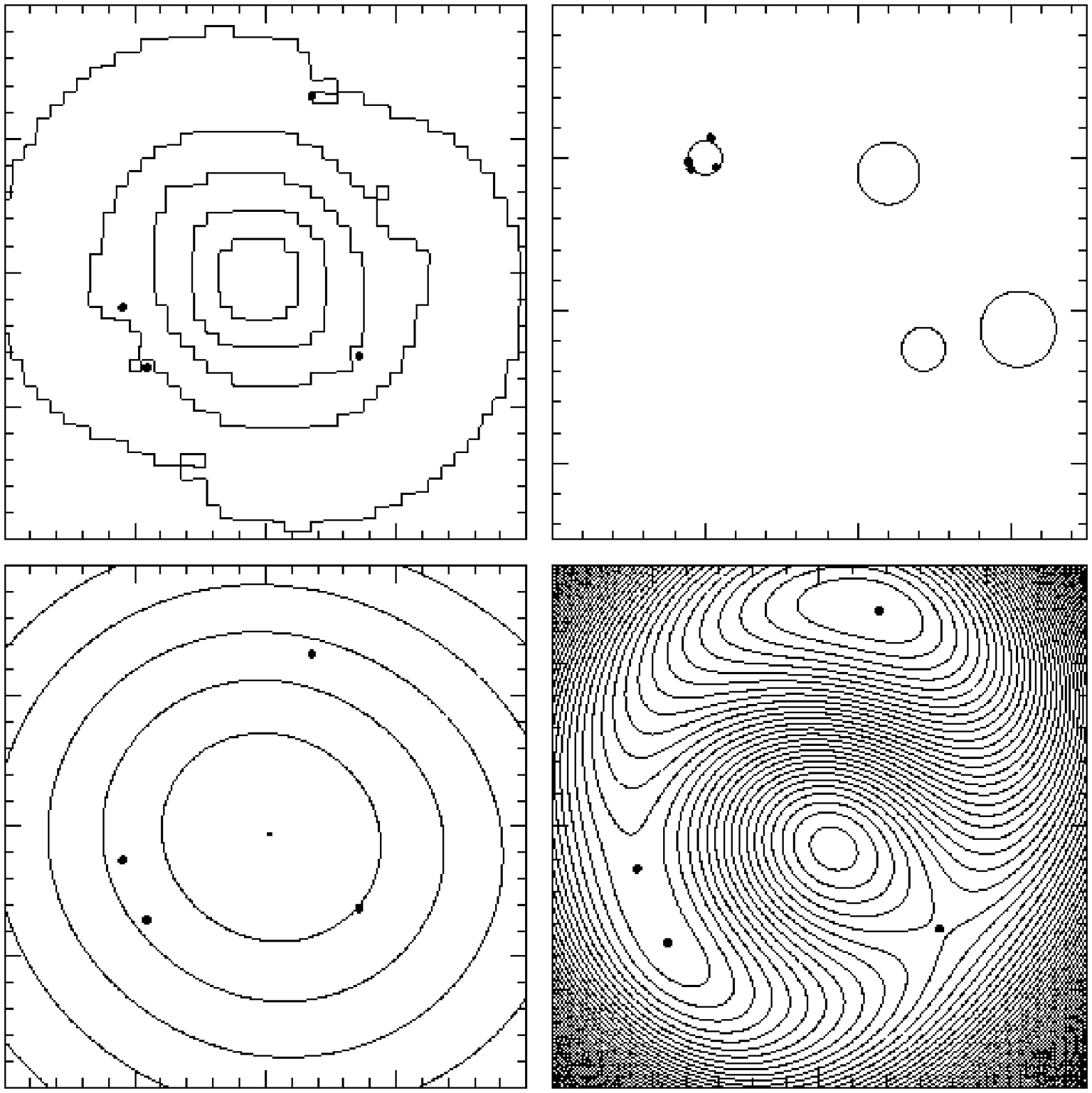,\hsize,\hsize] 3. A reconstruction constrained by
the image positions from \Kri\ and the time delays from \BK's
re-analysis of \Schech, and by $h=0.42$.
\endgraf
The four panels follow the plan of Figure~2, except that in the upper
right panel the galaxy masses are indicated by formal Einstein rings
(see equation \\{formal_Ering}).

\fig[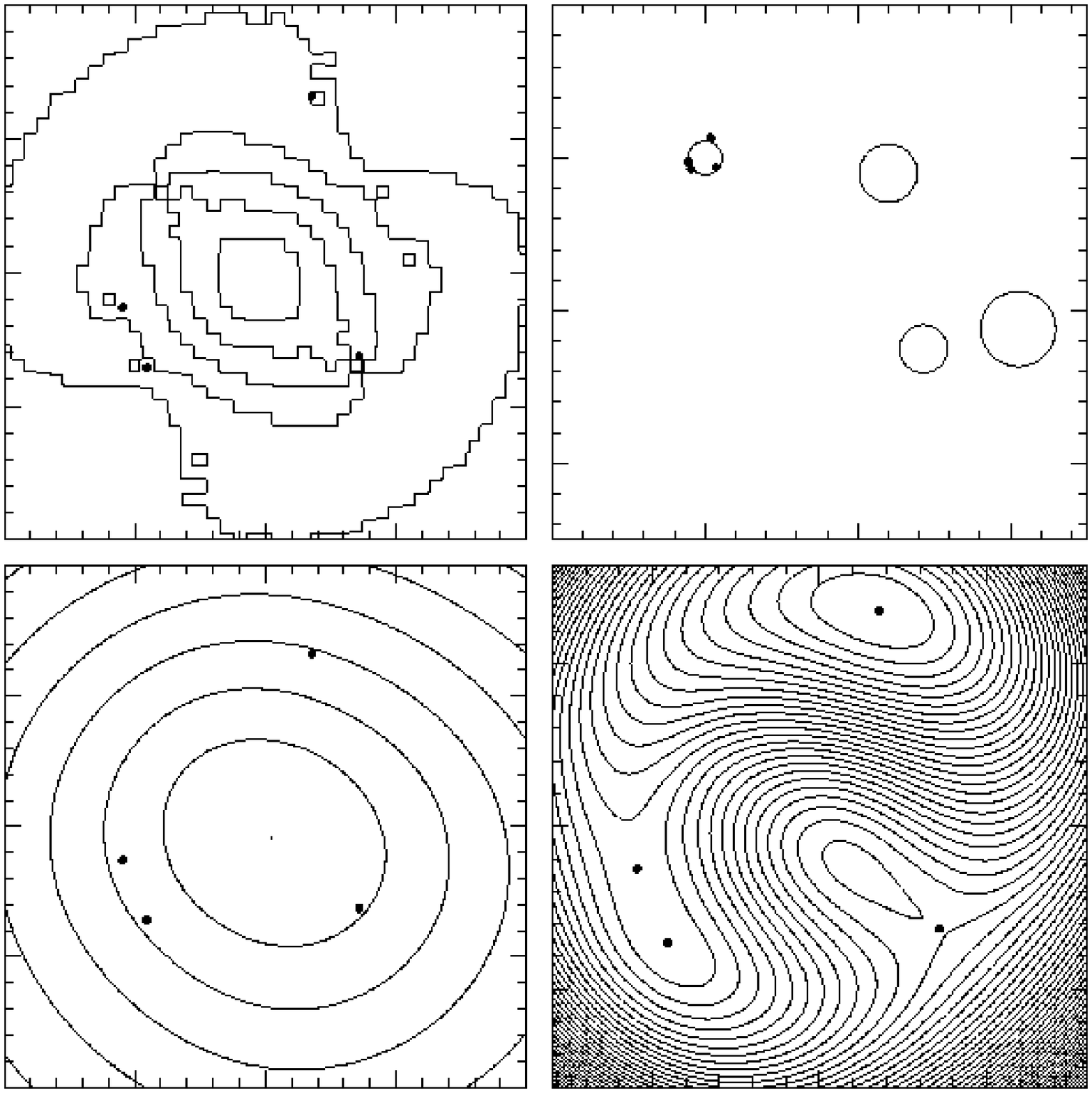,\hsize,\hsize] 4. Like Figure~3, except that
$h=0.63$.

\fig[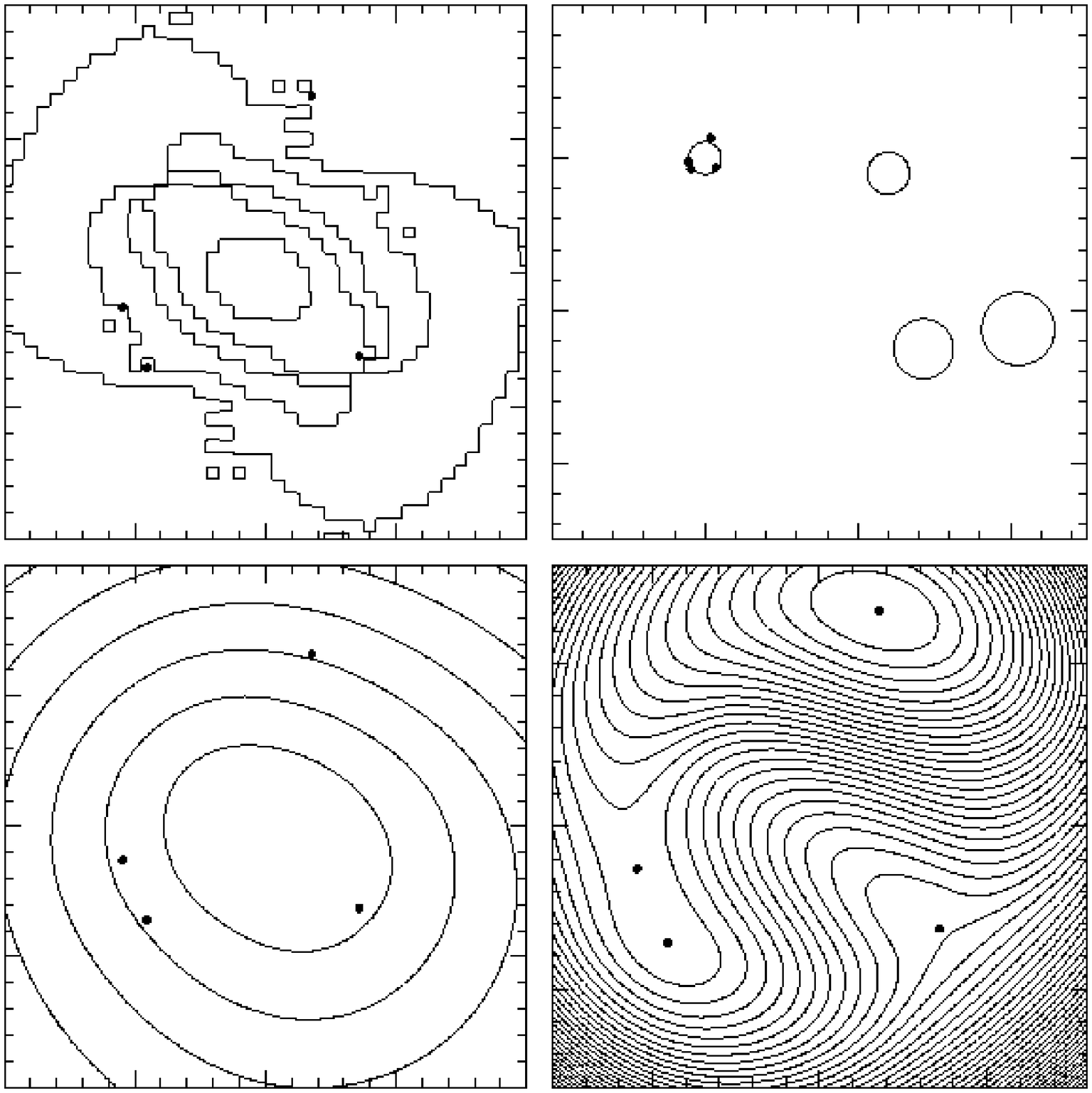,\hsize,\hsize] 5. Like Figures~3 and 4, except
now $h=0.84$. 

Now we discuss the reconstructions in Figures~3--5, which form the
main results of this paper.  These use the time delays $t_{\rm
BC}=25.0\;{\rm days}$ and $t_{\rm AC}/t_{\rm BA}=1.13$ from \BK's
re-analysis of the \Schech\ data, and include the other group galaxies
as point masses. On the basis of
the magnitudes in \Kri\ we loosely constrained the allowed
mass range of these galaxies (usually known as G1, G2, G3) relative to
the main lensing galaxy G as follows: (i)~The total mass of G1, G2 and
G3 is no more than 10 times that of G; (ii)~each of G1, G2 and G3 has
more mass than G; (iii)~G1 has more mass than G2 and G3 put together;
(iv)~G2 and G3 are at most a factor of 2 apart in mass. The masses of
the galaxies, one of the outputs of the code, are proportional to
the area of the circle in the diagrams. 

As we already pointed out, realistic galaxy profiles are easy to obtain
with our method,
hence $h$ is not usefully constrained by the current observations in
this lens system. Since this is the case, we decided to carry out mass 
reconstructions for three plausible values of $h$.
Figure~3 is constrained by $h=0.42$, Figure~4 by $h=0.63$ and 
Figure~5 by $h=0.84$. We chose these values to coincide
with the range found by \Schech\ in their modelling. 

The mass distribution looks different in the three cases.
For $h=0.42$, the galaxy is reconstructed as a possible early-type,
or a face-on spiral. For $h=0.84$, the galaxy is rather elongated, 
and so could be an edge-on disc, or highly flattened elliptical. 
Note that in all three cases, the major axis is aligned with the 
major axis of the group. The effect is barely noticeable for a low 
Hubble constant (see outer isophotes in Figure 3), and is most 
pronounced for a high Hubble constant. Is this alignment physically 
motivated? It is generally observed that brightest cluster galaxies 
(BCGs) in clusters are aligned with clusters' major axis
(Trevese {\it et al.} 1992, Fong {\it et al.} 1990). Similar 
observational evidence for poor clusters and groups
is less conclusive; Mendes de Oliveira \& Hickson (1994) based on
a sample of $\sim 100$ Hickson's compact groups show that BCGs are
not preferentially aligned with groups' axis. However, in our case
it is the alignment of a non-BCG with the group's major axis that 
is of interest,
and to our knowledge there is no observational data in this regard.
Ciotti \& Dutta (1994) performed numerical simulations to determine
whether clusters' tidal field can produce alignment in elliptical
galaxies located outside clusters' core. They find that an
initially elliptical galaxy gets aligned with clusters' major axis
due to the tidal field of a spherical cluster.
However, it is still unclear if such an alignment would take place
in a galaxy group. We remark that the boxy appearance of
our galaxy mass maps are not entirely dissimilar from 
the isodensity profiles obtained by Ciotti \& Dutta (see their 
Figures 8 and 9). Thus the alignment of the major axes of the lensing 
galaxy's major and its parent group are not necessarily unphysical.

Figures~3--5 indicate that higher values of $h$ require more elliptical
mass distribution in the lensing galaxy. This can be intuitively 
understood as follows. The time delay between any two images, 
$[(1+z_{\rm L}) {\Dl\Ds/c\Dls}]\cdot
[\tau(\btheta_1)-\tau(\btheta_2)]$, where $\tau(\btheta)$
is given by equation (\\{delayeqn}), is inversely proportional to $h$. 
Since time delays are fixed by observations, a higher value of $h$ 
necessitates the scaled time delays, 
$\tau(\btheta_1)-\tau(\btheta_2)$ to be proportionately 
larger. The scaled time delays between roughly equally spaced
(quadruple) images tend to be larger for elliptical lenses:
The time delay surface consists of a geometrical and gravitational 
part (see equation [\\{delayeqn}]), that are added together. The 
images are formed at stationary points of the resulting surface, and 
the time delay between them is just the difference in the `height' of 
the time delay surface at the image points. Then one can see that in 
general, when the gravitational potential of an elliptical lens is 
added to the geometrical part, the resulting stationary points will 
lie at more varied heights, implying larger scaled time delays, compared 
to a case with a circularly symmetric lens. Therefore, more elliptical 
lenses require larger $h$, if the observed time delays are fixed.

%% file: conc.tex
\section 5. Conclusions

In a multiple-image QSO, the time delay between two images
can be written as
$$ H_0^{-1} D(z,q_0) \Fl $$
where $D(z,q_0)$ is a dimensionless factor depending on the source and
lens redshifts and (weakly) on $q_0$, and $\Fl$ is of order the image
separation but its exact value depends on the distribution of the
lensing mass.  The uncertainty in $H_0$ estimates
from lensing comes almost entirely from the uncertainty in $\Fl$.
The recent measurement by \Schech\ of time delays in \PG\ are an
important advance, because positions of four images, rather than two,
and time delay measurements between two pairs of images, rather than
one pair help constrain $\Fl$ much better than in two-image
systems, like Q0957+561 A\&B.

Several authors have published parametric model fits to the \pg\
system; we may roughly summarize their experience as follows.
\item{(i)} A simple model suffices for a rough fit to the image
positions, but to the best available astrometry one needs external
shear (such as from the other group galaxies).
\item{(ii)} Fitting the time delay ratio $t_{{\!}\rm AC}/t_{{\!}\rm BA}$
is much more difficult; fits vary from marginally consistent to
inconsistent.
\item{(iii)} Different models give very different values of $H_0$. In
particular, the model fits in \Schech\ give $H_0$ ranging from 42 to
84\thinspace\kmsmpc, though the lower value is argued as more
plausible.
\par\noindent These points prompted us to think that maybe the
parametric models tried so far are only exploring a small part of
`model space', to know for sure we would have to try non-parametric
models.

We have therefore developed a non-parametric or free-form lens
modelling technique, which turns out to be rather simple to do. Image
positions and time delay ratios constrain a pixellated mass
distribution through linear equations.  Assumptions like $180^\circ$
rotation symmetry and density gradient always pointing inwards are
also readily cast as linear constraints.  Implementing all this, we
find that it is almost embarrassingly easy to fit the lensing data for
any plausible $H_0$ with plausible looking models.

We then try to pose a somewhat different question: what do different
ranges of $H_0$ imply about the lensing galaxy that might possibly be
distinguished by detailed observations of it?  (At present only a
centroid and a circular model for the galaxy light profile are
available, from \Kri.)  Some indications might be given by examining
mass models that, while satisfying the lensing data, follow the light
as closely as possible.  We construct such models for $h=0.42$, 0.63
and 0.84.  The $h=0.42$ model has a mildly elliptical mass profile
with some twist in the isodensity contours.  The $h=0.63$ model has
significant ellipticity inwards of the images, and the $h=0.84$ still
more. In both cases the surface density rises sharply very near
image~C and then flattens.  Such a system could be a late elliptical,
but a nearly edge-on disc would be more likely.  That the $h=0.63$ and
$h=0.84$ models have their flattening aligned with the group direction
may or may not be a physical feature of some outlying members of
galaxy clusters and groups.  
Another possibility is that the lensing galaxy is a
merging system, and for $h=0.63$ and 0.84 our density gradient
condition has forced the reconstruction into an awkward shape. 

In summary, the value of $H_0$ derived from \pg\ system and the
morphology of the main lensing galaxy are degenerate.  A round lensing
galaxy implies low $H_0$, high ($>\rm E5$) ellipticity oriented with
the group direction implies high $H_0$.

%% file: refs.tex
\parindent=0pt \frenchspacing
\everypar{\hangindent 20pt \hangafter 1}
 
\section References

Angonin-Willaime M.-C., Hammer, F., \& Rigaut, F. 1993, in 
{\it Gravitational Lenses in the Universe}, eds. J. Surdej,
D. Fraipont-Caro, E. Gosset, S. Refsdal, \& M. Remy
(Li\`ege: Institut d'Astrophysique), p. 85

Bar-Kana, R. 1997, Preprint,
also available as astro-ph/9701068 (\BK)

Blandford, R., \& Narayan, R. 1986, ApJ, 310, 568

Ciotti, L., \& Dutta, S. N. 1994, MNRAS, 270, 390

Christian, C. A., Crabtree, D., \& Waddel, P. 1987, ApJ, 312, 45

Courbin F., Magain, P., Keeton, C. R., Kochanek, C. S., Vanderriest, C.,
Jaunsen, A. O., \& Hjorth, J. 1997, A\&A Letters, in press, also
available as astro-ph/9705093

Fong R., Stevenson P. R. F., \& Shanks, T. 1990, MNRAS, 242, 146

Foy, R., Bonneau, D., \& Blazit, A. 1985, A\&A,  149, L13

Green, R. F., Schmidt, M., \& Liebert, J. W. 1986, ApJS, 61, 305

Hege, E. K., {\it et al.} 1981, ApJL 248, L1

Henry, J. P., \& Heasley, J. N. 1986, Nature, 321, 139

Jacoby, G. H., {\it et al.} 1992, PASP, 104, 599

Keeton, C. R., \& Kochanek, C. S. 1997, Preprint, 
also available as astro-ph/9611216

Kochanek, C. S. 1991, ApJ, 373, 354

Kristian, J. {\it et al.} 1993, AJ, 106, 1330 (\Kri)

Kundi\'c, T., Cohen, J. G., Blandford, R., D., \& L. M. Lubin. 
1997, Preprint, also available as astro-ph/9704109

Mendes de Oliveira, C., \& Hickson, P. 1994, ApJ, 427, 684

Michalitsianos, A. G., Oliversen, R. J., \& Nichols, J. (1996) 
ApJ, 461, 593

Narasimha, D., Subramanian, K., \& Chitre, S. M. 1982, MNRAS, 200, 941

Schechter, P. L., {\it et al.} 1997, ApJL, 475, L85 (\Schech)

Shaklan, S. B., \& Hege, E. K. 1986, ApJ, 303, 605

Tonry, J. L. 1997, Preprint, also available as astro-ph/9706199

Trevese D., Cirimele G., \& Flin, P. 1992, AJ, 104, 935

Tripp, T. M., Green, R. F., \& Bechtold, J. 1990, ApJL, 364, L29 

Vanderriest, C., Vlerick, G., Lelievre, G., Schneider, J., Sol, H., 
Horville, D., Renard, L., \& Servan, B.  1986, A\&A, 158, L5

Weymann, R. J., Latham, D., Angel, J. R. P., Green, R. F., Liebert, 
J. W., Turnshek, D. A., Turnshek, D. E., \& Tyson, J. A. 
1980, Nature, 285, 641

Witt, H.-J., \& Mao, S. 1997, Preprint,
also available as astro-ph/9702021

Young, P., Deverill, R. S., Gunn, J. E., Westphal, J. A., \&
Kristian, J. 1981, ApJ, 244, 723

%% file: ack.tex
\section Acknowledgments

LLRW would like to acknowledge the support of the PPARC
fellowship at the Institute of Astronomy, Cambridge, UK.

%% file: app.tex
\eqnum=0
\def\theeqnum{\rm A.\the\eqnum}

\section Appendix

This short appendix is about the integrals (\\{pixelintegs}) over
square pixels to give the contribution of one pixel to the
lens potential and bending angles at arbitrary $\btheta$.

Below we will freely use cartesian or polar forms for the two
dimensional angle $\btheta$.

To work out the definite integrals (\\{pixelintegs}), it is convenient
to first work out the indefinite integrals, which are
$$ \eqalign{ & ~\psi = {1\over2\pi}
       \Big( x^2\arctan(y/x) + y^2\arctan(x/y)
              + xy (2\ln r-3) \Big),  \cr
             & ~\alpha_x = {1\over\pi}
       \Big( x\arctan(y/x) + y\ln r \Big), \cr
             & ~\alpha_y = {1\over\pi}
       \Big( y\arctan(x/y) + x\ln r \Big). \cr }
   \putnum$$
The arctans are defined to range from $-\pi$ to $\pi$. We are
free to add arbitrary functions $f(x)$ and $g(y)$ to any of the
indefinite integrals (\prevenum1).  For example $2\pi~\psi=
(x^2-y^2)\arctan(y/x)+xy\left(2\ln r-3\right)$ is also a
legitimate indefinite integral; but it is discontinuous across the $x$
axis, and would give wrong answers if used below in (\prevenum0).

To get the definite integrals, we take the appropriate differences of
the pixel-corner values of the indefinite integrals.  Thus, if the
$n$th pixel is centred at $(x_n,y_n)$ we have:
$$ \eqalign{ & \psi_n(x,y) = ~\psi(x_+,y_+) + ~\psi(x_-,y_-)
                       - ~\psi(x_-,y_+) - ~\psi(x_+,y_-),\cr
   & x_\pm = x - x_n \pm a/2, \qquad y_\pm = y - y_n \pm a/2,\cr
  }  \putnum$$
and similarly for $\balpha_n$.